\documentclass[useAMS,usenatbib,usegraphicx]{mn2e}

\usepackage{amsmath}
\usepackage{amssymb}
\usepackage{graphicx}
\usepackage[dvipsnames]{xcolor}
\usepackage{subfigure}

\begin{document}

\voffset=-0.8in

\title[Accretion rate onto IMBHs]{Numerical estimates of the accretion rate onto
intermediate-mass black holes}
\author[C. Pepe et al.]{C. Pepe$^{1,2}$\thanks{E-mail:
carolinap@iafe.uba.ar} and L. J. Pellizza$^{1,2}$\\
$^{1}$Instituto de Astronom\'ia y F\'isica del Espacio, Casilla de Correos 67,
Suc. 28, 1428, Buenos Aires, Argentina\\
$^{2}$Consejo Nacional de Investigaciones Cient\'ificas y T\'ecnicas, CONICET,
Argentina\\}

\date{Accepted 2013 January 11. Received 2013 January 11; in original form 2012 November 21}

\pagerange{\pageref{firstpage}--\pageref{lastpage}} \pubyear{2012}

\maketitle

\label{firstpage}

\begin{abstract}
The existence of intermediate-mass ($\sim 10^3 M_\odot$) black holes in
the center of globular clusters has been suggested by different observations.
The X-ray sources observed in NGC~6388 and in G1 in M31 could be
interpreted as being powered by the accretion of matter onto such objects. 
In this work we
explore a scenario in which the black hole accretes from the cluster 
interstellar medium, which is generated by the mass loss of the red giants in the cluster.
We estimate the accretion rate onto the black hole and compare it to the values
obtained via the traditional Bondi-Hoyle model. Our results show that the accretion
rate is no longer solely defined by the black hole mass and the ambient parameters but also
by the host cluster itself. Furthermore, we find that the more massive globular clusters with large
stellar velocity dispersion are the best candidates in which accretion onto IMBHs could be detected. 

\end{abstract}


\section{Introduction}

The existence of intermediate-mass black holes ($10^2$--$10^4 M_{\odot}$, IMBHs)
was suggested, among other observations, by the detection of ultraluminous X-ray
sources in nearby galaxies \citep[see][for a review]{Feng2011}. The X-ray fluxes
and the distances measured for these sources imply luminosities (assuming
isotropic emission) $L > 10^{39}\ {\rm erg\ s}^{-1}$, in excess of the Eddington
luminosity for an accreting stellar-mass compact object ($M \sim 10 M_{\odot}$).
The simplest explanation for these high luminosities is the presence of
accreting objects with masses in the range of those of IMBHs, but other
hypotheses have been proposed. \citet{Koerding02} suggested that the emission of
ULXs could be beamed, hence implying lower luminosities for these sources, while
\citet{King01} proposed that ULXs emit indeed in a super-Eddington regime with
mild geometrical collimation of their photon emission (a factor $\lesssim10$) due to outflows.
Although in these alternative models accreting stellar-mass black holes could
explain the observed luminosities, they fail to account for sources with
$L \gtrsim 10^{41}\ {\rm erg\ s}^{-1}$, which still require the accreting compact object
to be an IMBH \citep{Feng2011}.

On the other hand, different theoretical models for the origin of IMBHs have
been proposed. According to \citet{Fryer2001}, IMBHs would be the dead fossils
of primordial (Population III) stars, while \citet{Miller2002} have shown that
black holes of $10^3 M_\odot$ could form in globular clusters (GCs) as a
consecuence of the merger of stellar-mass black holes. A similar mechanism was
proposed by \citet{PortegiesZwart2004}, who have shown that a runaway collision
of massive stars in a GC results in an IMBH at its centre. Hence, GCs have
become the main candidates to host these objects. The extrapolation of the
relation between the central black hole mass and the bulge mass of galaxies
 \citep[e.g.,][]{Magorrian1998} also points to the existence of IMBHs in GCs.

Following the predictions of theoretical models, several attempts to detect
IMBHs at the centres of GCs were done. Stellar density profiles and stellar
dynamics measurements in the central regions of some GCs suggest the existence
of central objects with masses $\sim 10^3 M_{\odot}$. \citet{Miocchi2007}
developed a model for the stellar distribution in GCs, including the effects of
an IMBH. Those models containing an IMBH yield results consistent with the 
surface brightness and velocity dispersion
profiles obtained by \cite{Noyola2006} for the Galactic GCs NGC~2808, NGC~6388, M80, M13, M62 and
M54, and also G1 in M31. \citet{VandenBosch2006}
constructed dynamical models of M15 and estimated the central mass in
$3400\ M_\odot$. However, the nature of this concentrated mass can not be
distinguished: it could be either an IMBH or a large number of stellar-mass
compact objects, or a combination of both. On the other hand,
\citet{McLaughlin2006} fitted the proper motion dispersion profile of 47 Tuc,
obtaining an estimate of its central point mass of $1000-1500\ M_\odot$ at 68\%
confidence level. \citet{Noyola2010} analyzed the surface brightness and
velocity dispersion profiles of $\omega$~Cen. Both profiles show a central cusp,
a 4000$M_{\odot}$ IMBH being the best explanation for these observations.
\citet{Gebhardt2002} reported a $2\times 10^4 M_{\odot}$ point mass at the center
of G1, based on photometric and kinematical data in the central areas.
However, other authors arrived at different conclusions.
For example, \cite{Vandermarel2010}  and \cite{Baumgardt2003} constructed
dynamical models for $\omega$~Cen and G1, respectively, claiming that the
presence of an IMBH is not needed to fit the available data. Because of this
lack of consensus, the study of IMBHs is still an open issue and efforts need
to be made in order to clarify the subject. Moreover, although
kinematical and photometric data are suitable to show the presence of central
mass concentrations in GCs, at present they are unable to determine the nature
of these concentrations, namely if they comprise a single massive object or a
collection of stellar-mass remnants. Complementary data are needed to prove any
of these hypotheses.

As in other systems containing compact objects, the detection of ongoing
accretion may help to place constraints on the existence and properties of
IMBHs. Hence, searches of central X-ray sources have been performed in several
GCs. The detection of central sources with properties consistent with those
expected from an IMBH has been reported only for two clusters: NGC~6388 and G1
\citep{Nucita2008,Kong2010}. In other clusters, only upper limits for the
X-ray luminosity of a central source have been obtained, constraining it to be
lower than $\sim 10^{31-32}\ {\rm erg\ s^{-1}}$ \citep[and refereces
therein]{Maccarone2004}. Assuming
that the IMBH accretes from the intracluster medium (ICM), that the ICM has a
density similar to that measured by \citet{Freire2001} in NGC~104, and that the
radiative efficiency is similar to that observed in other black hole systems,
\citet{Maccarone2004} concludes that the accretion rate must be far lower than
that predicted by the standard Bondi-Hoyle model \citep{Bondi1944} in order
to match the non-detections of central sources. Based on the
correlation between X-ray and radio luminosity of accreting stellar-mass black
holes \citep{Gallo2003}, this author also argues that the radio emission of the
system would be more easily detectable than the X-ray emission. However, no
central radio source has been detected in Galactic GCs which, according to
\citet{Maccarone2004}, imposes strong constraints to the masses of the
hypothetical IMBHs. Only \citet{Ulvestad2007} reported the detection of a radio
source at about 1~arcsec from the centre of G1 in M31. They concluded that this
radio emission is consistent with that expected for a $2 \times 10^4\ M_\odot$
IMBH accreting at the center of the cluster. Nevertheless, this result has been
challenged by recent observations \citep{MillerJones2012}.

It is clear that the detection of (or the failure to detect) accretion-powered
sources in the centres of GCs is crucial for the investigation of the existence
and nature of IMBHs. In this context, the interpretation of both X-ray and radio
data have been done in the past using simple models. In particular, Bondi-Hoyle
accretion has been used, which assumes a point mass (the IMBH) accreting from a
homogeneous, static medium. In a previous work \citep{Pepe2012} we studied the
accretion of dark matter onto IMBHs. We took into account the gravitational pull
of the cluster on the accreted matter and found that the accretion rate depends
on the cluster mass, unlike the Bondi-Hoyle accretion rate that scales as the
square of the IMBH mass. The Bondi-Hoyle accretion rate is retrieved only for
ultrarelativistic fluids, which are not influenced by the presence of the
cluster. As the ICM is a non-relativistic fluid, we expect that its behaviour
departs from that predicted by the standard Bondi-Hoyle theory. Moreover, the
ICM of a GC is not a homogeneous, static medium, because it is fed by the
red giant stars and cleansed by the passages of the cluster through the Galactic
plane \citep{Roberts88}. The influence of these effects on the accretion rate
should be quantified to make a proper comparison between models and X-ray or
radio observations.

In this work we develop a simple numerical model for the accretion flow of the
ICM onto an IMBH at the centre of a GC, including the cluster gravitational pull
and the ICM sources. We estimate the accretion rate and its dependence on both
the IMBH mass, and the GC and ICM properties, and compare them with those
predicted by the standard Bondi-Hoyle model. In Sect.~\ref{modelo} we describe
the hydrodynamical equations of the flow and discuss its main characteristics,
while in Sect.~\ref{resultados} we solve these equations for Milky Way GCs with
different masses of the IMBHs and temperatures of the ICM, stressing the
differences between our results and those obtained with the Bondi-Hoyle model.
Finally, in section \ref{discusion} we discuss the consequences of our results
on the search for IMBHs, and present our conclusions.

\section{The model}
\label{modelo}

\subsection{Hydrodynamical equations}

In order to understand the accretion process by an IMBH in a GC, we study the
dynamics of the ICM in the presence of the cluster plus IMBH gravitational
potential. We aim at computing the accretion rate, to compare it with the
standard Bondi-Hoyle theory. The ICM is generated by the mass loss of the red
giants of the cluster. Assuming that the distribution of these stars follows
the stellar mass density of the cluster, and that the average mass loss rate is
the same for all red giants, the rate at which the density of the ICM
increases due to the injection of matter by these stars, at any position within the cluster can be written as

\begin{equation}
 \dot{\rho} = \alpha \rho^*
\end{equation}

\noindent where $\rho^*$ is the stellar mass density. The right hand
side describes the gas injection by the stars at a fractional rate 
$\alpha$, for which theoretical and observational works obtain estimations in the range
$10^{-14}$--$10^{-11}\ {\rm yr}^{-1}$ \citep[and references therein]{Fusci1975,Scott1975,Dupree1994, Mauas2006}.
The cluster plus IMBH gravitational field can be described by the model
developed by \citet{Miocchi2007}. This model is basically a \citet{King1966}
model with the presence of a black hole at the centre of the cluster, which
modifies the dynamics of the stars in the inner region. Following
\citet{Scott1975}, we assume that the ICM is an ideal gas, and that its flow can
be considered steady, spherically symmetric and isothermal. Under these
hypotheses, the flow is governed by continuity and Euler's equations. The former
is

\begin{equation}
\frac{1}{r^2}\frac{d}{dr}(\rho r^2 u)= \alpha \rho^*,
\label{cont}
\end{equation}

\noindent
where $r$ is the radial coordinate and $u$ and $\rho$ the velocity and density of
the flow, respectively. Euler's equation is

\begin{equation}
\rho u \frac{du}{dr}= -\frac{k_{\mathrm B} T}{\mu}\frac{d\rho}{dr} - \frac{G M(r)
\rho}{r^2}-\alpha u \rho^*,
\label{euler}
\end{equation}

\noindent
where $G$ is the gravitational constant, $k_\mathrm{B}$ is Boltzmann's constant,
$\mu$ is the mean molecular mass of the ICM, and $M(r)$ the sum of the stellar
mass $M^*(r)$ inside radius $r$ and the central IMBH mass $M_{\rm BH}$. It is
assumed here that the material is injected with null velocity in the flow.

These equations can be simplified introducing the variable
$\tilde{q} = q / \alpha$, where $q = \rho u r^2$ is the bulk flow, giving

\begin{equation}
\label{caudal_prima}
\frac{d\tilde{q}}{dr} =  \rho^* r^2, 
\end{equation}

\begin{equation}
\label{u_prima_caudal}
\frac{du}{dr} =  \frac{u}{u^2 - c_\mathrm{s}^2}\left( \frac{2 c_\mathrm{s}^2}{r} -
\frac{G M(r)}{r^2} - \frac{(u^2 + c_\mathrm{s}^2) r^2\rho^*}{\tilde{q}}\right),
\end{equation}

\noindent
where $c_\mathrm{s} = k_\mathrm{B} T \mu^{-1}$ is the sound speed. It is important
to note that, with this definition, we can solve the equations independently of
$\alpha$. However, $\alpha$ must be known to calculate the density and the
accretion rate.

Equation \ref{caudal_prima} can be integrated, giving

\begin{equation}
\tilde{q} = \int^r_0 \rho^* r'^2 dr' + \tilde{q}_0 = \frac{M^*(r)}{4\pi} +
\tilde{q}_0,
\label{caudal}
\end{equation}

\noindent
where the integration constant $\tilde{q}_0$ is proportional to the accretion
rate of the black hole. Although the integration should be done from the
Schwarzschild radius, this radius is negligible with respect to all the scales
in our model, hence we take it as zero.

To perform the integration, we define adimensional variables $\xi = r r_0^{-1}$,
$\psi = u \sigma^{-1}$ (and, therefore, $\psi_{\rm s} = u_{\rm s} \sigma^{-1}$ ), $\omega = \tilde{q} (\rho_0 r_0^3)^{-1}$,
$\Omega^*(\xi) = M^*(r) (4 \pi \rho_0 r_0^3)^{-1}$,
$\Omega_{\rm BH}= M_{\rm BH} (4 \pi \rho_0 r_0^3)^{-1}$, and
$\Omega(\xi) = \Omega^*(\xi) + \Omega_{\rm BH}$, where $r_0$ is the King radius,
$\rho_0$ is the cluster central density, and
$\sigma^2 = 4 \pi G \rho_0 r_0^2 / 9$ is the velocity dispersion parameter. With
these definitions, Eqns.~\ref{u_prima_caudal} and \ref{caudal} can be rewritten
as

\begin{equation}
\label{omega}
\omega = \Omega^*(\xi) + \omega_0,
\end{equation}

\begin{equation}
\label{psi_prima_ad}
\frac{d\psi}{d\xi}=\frac{\psi}{\psi^2 - \psi^2_{\rm s}} \left(\frac{2\psi^2_{\rm s}}{\xi} - 
\frac{d\omega}{d\xi}\frac{(\psi^2_{\rm s} + \psi^2)}{\omega} -
\frac{9 \Omega(\xi)}{\xi^2} \right).
\end{equation}

\noindent
These equations describe the flow dynamics in our model. They represent an
extension of the Bondi-Hoyle expressions to the case in which the fluid has
sources, and a gravitational field other than that of the accretor acts on it.
These effects are represented by the second and the third terms in brackets in
the right-hand side of Eqn.~\ref{psi_prima_ad}, and the first term in the
right-hand side of Eqn.~\ref{omega}. The boundary conditions and integration
method for this set of equations are described in the following section.

\subsection{Boundary conditions} 

In order to integrate equation \ref{psi_prima_ad}, boundary conditions must be
set. Far away from the acretting black hole, the velocity of the flow must be
positive as there is no matter source outside the cluster. On the other hand, on
the black hole surface, matter can only fall inwards as there is no pressure
gradient that supports the gravitational pull of the black hole and the cluster.
As a direct consecuence, there exists a stagnation radius $\xi_{\rm st}$ at which
$u = 0$. Evaluating Eqn.~\ref{omega} at the stagnation radius gives

\begin{equation}
\Omega^*(\xi_{\rm st}) = -\omega_0,
\label{omega_acc}
\end{equation}

\noindent
which means that all the matter ejected by the red giants inside $\xi_{\rm st}$ is
acretted by the IMBH. This is a direct consequence of the hypothesis of
stationarity. Thus, the stagnation radius separates two distinct regions in
space: a regime of accretion develops in the inner region, and a wind solution
exists in the outer one. It is important to highlight that, so far, the
stagnation radius cannot be uniquely defined by these equations. For every
stagnation radius there exist a solution to the equations. However, they are not
all physically acceptable, as the Rankine-Hugoniot conditions for the continuity
of the flow at the stagnation radius require that densities at both sides of
$\xi_{\rm st}$ must be equal.

As in the Bondi-Hoyle problem \citep[e.g.,][]{Frank02}, it can be seen from
Eqn.~\ref{psi_prima_ad} that there exist singular values $\xi_{\rm s}$ (called
sonic radii) for which either the velocity equals the sound speed in the medium
or the aceleration of the flow is null. This is the case when the expression in
brackets in Eqn.~\ref{psi_prima_ad} cancels, 

\begin{equation}
\label{funcion_sonico}
2\psi^2_{\rm s} \left( \frac{1}{\xi} - \frac{1}{\omega}\frac{d\omega}{d\xi}\right)
- \frac{9 \Omega(\xi)}{\xi^2} = 0.
\end{equation}

\noindent
For the same reasons of the classical analysis of the Bondi-Hoyle problem, only
the transonic curves are consistent with the properties of an accretion flow. 

Given that $\xi_{\rm st}$ is not known a priori, the problem was solved
numerically for a grid of values of $\xi_{\rm st}$ between 0 and the cluster tidal
radius. For each one of these values, sonic radii in the inner and outer regions
were searched for, using a bisection scheme to find the roots of
Eqn.~\ref{funcion_sonico}. Then, Eqn.~\ref{psi_prima_ad} was integrated (via a
fourth order Runge-Kutta scheme) inwards from the outer sonic point and outwards
from the inner one, up  to the stagnation radius. The difference between the
densities at each side of the stagnation point were calculated for each
$\xi_{\rm st}$, and interpolated to zero to find the true stagnation radius. Once
this value was found, we integrated Eqns.~\ref{omega} and \ref{psi_prima_ad} to
obtain the true density and velocity profiles. 

\begin{figure*}
\vskip 0.7cm
\includegraphics[width=0.95\columnwidth]{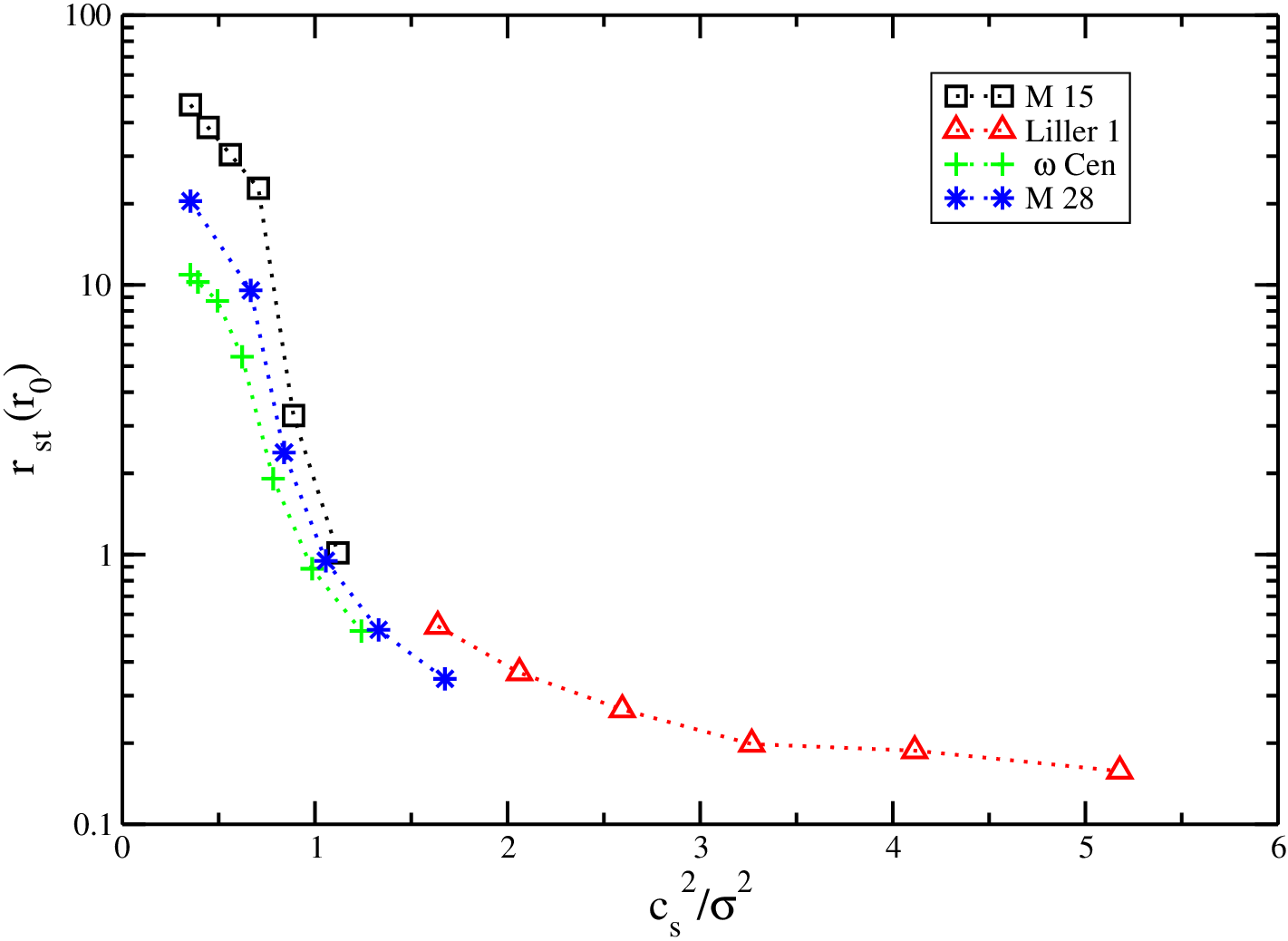}
\hskip 1cm
\includegraphics[width=0.95\columnwidth]{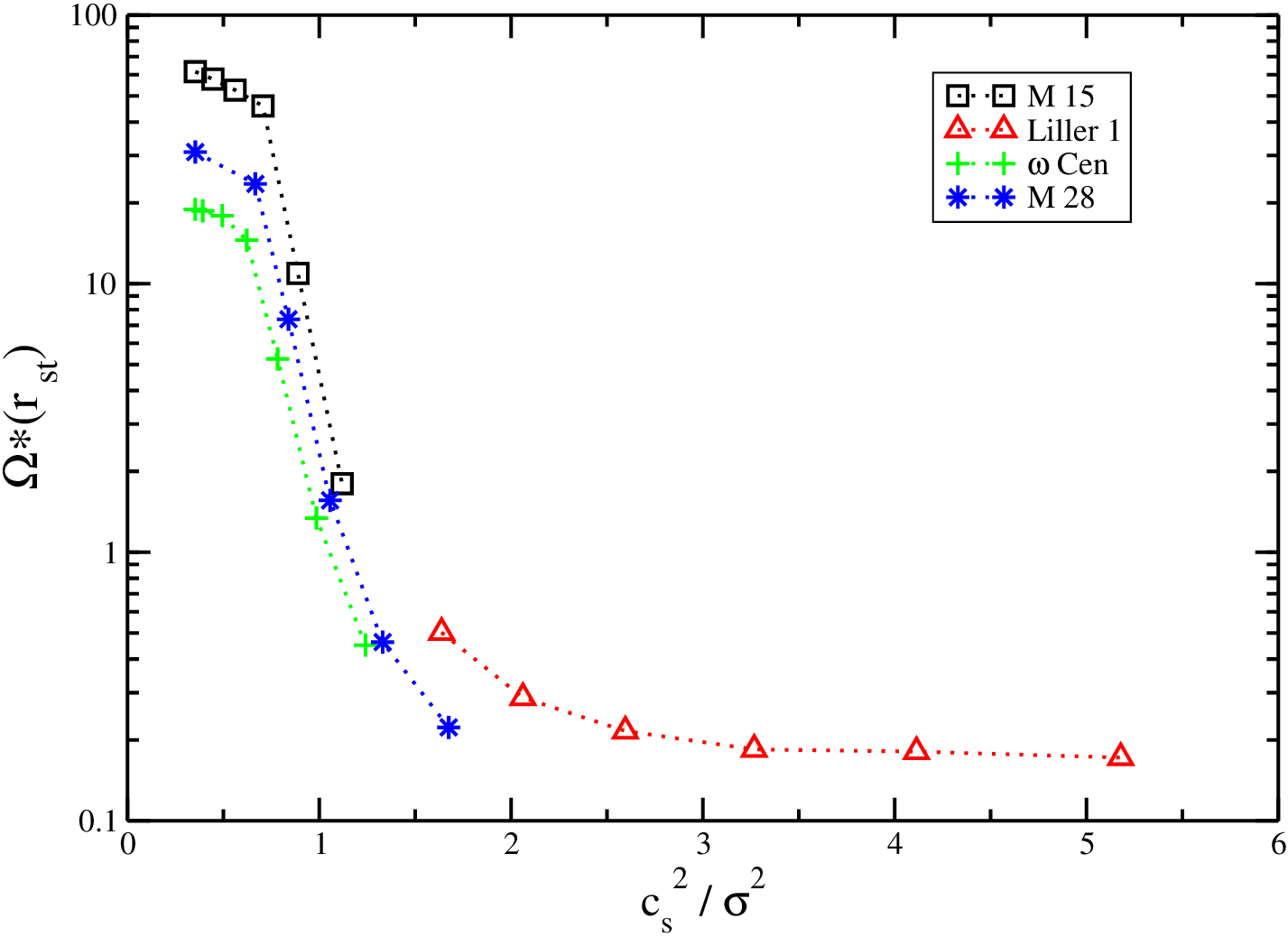}
\caption{Stagnation radius and accretion rate vs gas temperature for different clusters (squares for
M 15, triangles for 
Liller 1, plus signs for $\omega$ Cen and stars for M 28). The adimensional black 
hole mass $\Omega_{\rm BH}$ is 
$\sim$0.007 in all cases. A decreasing profile arises and the curves are placed in the graph ordered with increasing central 
cluster potential (from bottom to top).}
 \label{vstemp_adim}
\end{figure*}

\begin{figure*}
\vskip 0.7cm
 \includegraphics[width=0.95\columnwidth]{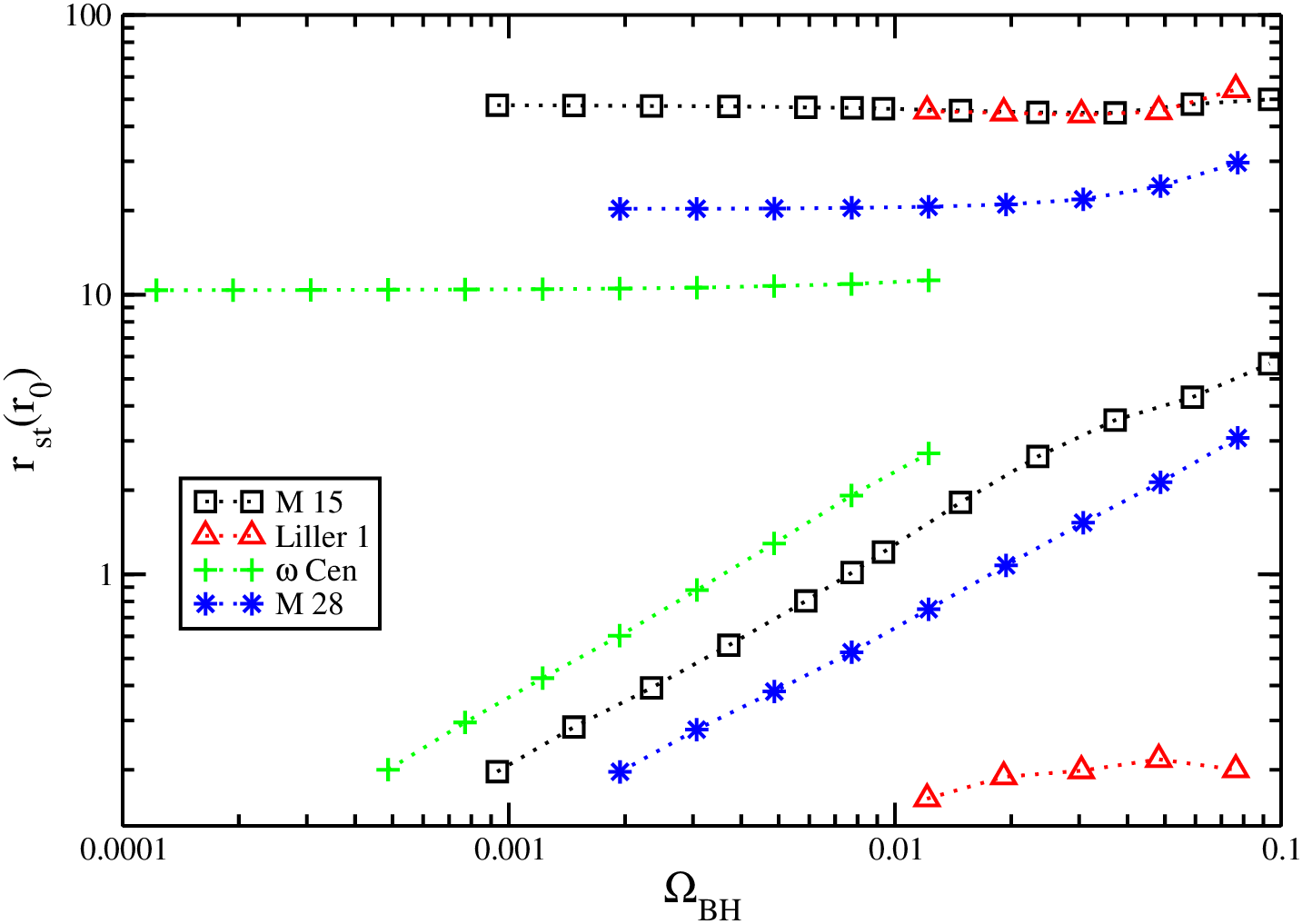}
\hskip 1cm
\includegraphics[width=0.95\columnwidth]{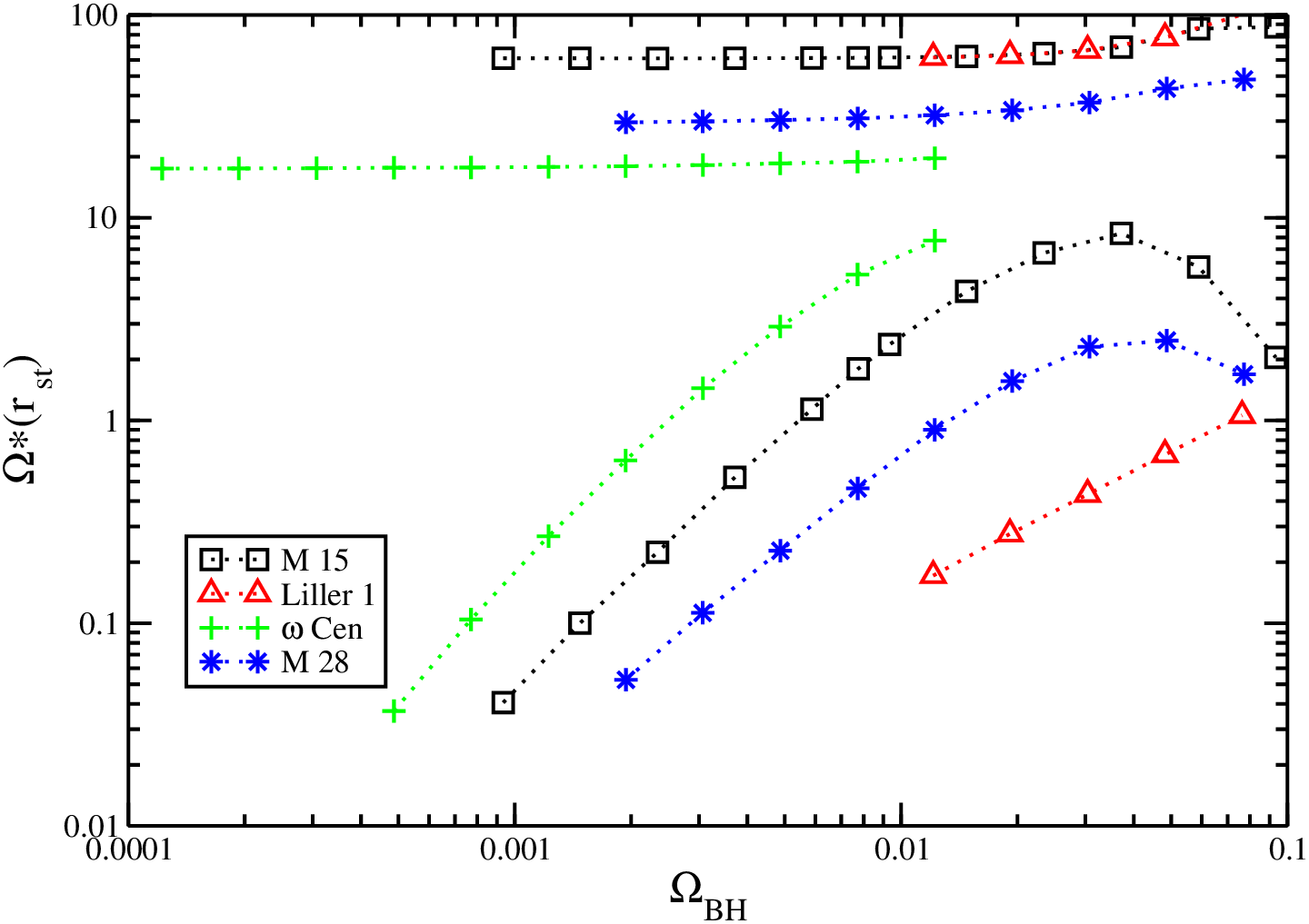}
\caption{Stagnation radius and accretion rate vs black hole mass  for different clusters (squares for M 15, 
triangles for 
Liller 1, plus signs for $\omega$ Cen and stars for M 28). The adimensional temperature $c_s^2 / \sigma^2$ is 
$\sim$0.34 for all the clusters in the HAR regime (upper set of curves). For those in a LAR regime (lower set of curves) the 
adimensional temperature is around 1, except for Liller 1, for which it is $\sim 3$. The curves are placed in the graph ordered with increasing central cluster potential 
(from bottom to top).
It can be seen that the dependence of the accretion 
rate on the black hole mass can be neglected compared to the temperature dependence. }
 \label{vsmasa_adim}
\end{figure*}

\section{Results}
\label{resultados}

\subsection{Individual clusters}

We first explore the behaviour of the accetion flow in a given GC. The cluster
is represented by the value of the concentration $c_{\rm GC}$ (defined as the
ratio of the cluster tidal radius to $r_0$), and two of the three parameters
$r_0$, $\rho_0$, $\sigma$. These parameters, together with the IMBH mass allow
us to construct the \citet{Miocchi2007} model for the stellar mass $M(r)$ of the
cluster uniquely. The concentration and IMBH mass define the shape of $M(r)$,
while the other parameters merely set its physical scale. Note that the same is
true for our hydrodynamical model, as the flow can be fully expressed using
non-dimensional variables defined in terms of those scaling parameters. Thus,
for a given GC there are only two free parameters in our model: the
non-dimensional temperature (or sound speed $c_{\rm s} / \sigma$) and IMBH
mass $\Omega_{\rm BH}$.

With the above considerations in mind, we took four sample GCs: NGC~7078 (M15),
Liller~1, NGC~6626 (M28) and NGC~5139 ($\omega$ Cen) , which span the concentration range of Milky Way
GCs (see Table~\ref{clusters}), and for each one we constructed different flow
models, performing a scan of the two free parameters. Although NGC~7078 and
Liller~1 have almost the same value for the concentration, they differ in the
scaling parameters, which allows us to extend the range of the non-dimensional
free parameters. The temperature covers the range $5000-15000\ {\rm K}$, which
is the range expected for the ICM \citep{Scott1975,Knapp1996,Priestley2011}. The IMBH
mass ranges from $10^2 M_\odot$ up to $10^4 M_\odot$. Heavier masses would produce
a strong effect on the stellar dynamics and structure of the cluster, which is
not observed, while for lower masses the assumption of the IMBH at rest at the
center of the cluster would not hold.

\begin{table}
\begin{tabular}{lcccc}
\hline
Cluster & $\sigma$ & $r_0$ & $\rho_0$ & $c_{\rm GC}$ \\
 & (${\rm km\ s}^{-1}$) & (${\rm pc}$) & ($M_\odot\ {\rm pc}^{-3}$) & \\
\hline
NGC 7078 (M 15) &10.8 & 0.42 & $1.12 \times 10^5$ & 2.29 \\
Liller 1   &5.05 & 0.15 &$1.90 \times 10^5 $& 2.3 \\
NGC 5139 ($\omega$ Cen) & 10.2& 3.58 & $1.41 \times 10^3$ & 1.31 \\
NGC 6626 (M 28) &7.8& 0.38 & $7.24 \times 10^4$ & 1.67 \\
\hline
\end{tabular}
\caption{Globular cluster parameters taken from \citet{Harris1996}. The central density was estimated assuming a 
mass-luminosity relation $M_{\odot} \sim L_{\odot}$.}
\label{clusters}
\end{table}

In Figs.~\ref{vstemp_adim} and \ref{vsmasa_adim} we show the stagnation radius
$\xi_{\rm st}$ and the (non-dimensional) accretion rate $\Omega(\xi_{\rm st})$ as a
function of the free parameters $c_{\rm s} / \sigma$ and $\Omega_{\rm BH}$,
respectively, with the other parameter kept fixed. It is worth pointing out that
the true accretion rate $\dot{M}$ can be recovered via
$\dot{M} = \alpha \Omega(\xi_{\rm st}) \rho_0 r_0^3$, with a suitable value for
$\alpha$ (see Sect.~\ref{discusion} for a discussion). It can be seen from the
left panel of Fig.~\ref{vstemp_adim} that the stagnation radius decreases with
the ratio $c_{\rm s} / \sigma$. This can be easily understood in terms of the
energetics of the gas: for higher temperatures the gas has more energy and can
escape from inner regions of the cluster, where the gravitational well is
deeper, hence moving the stagnation radius inwards. This results in a lower
accretion rate (right panel). It is interesting to note that the stagnation
radius curve steepens at $r_{\rm st} \sim r_0$, where $c_{\rm s} / \sigma \sim 1$.
The rapid increase of the stellar mass of the cluster at this radius results
then in an abrupt change in the accretion rate. This abrupt change suggests the
existence of two accretion regimes: at high temperatures
($c_{\rm s} / \sigma > 1$), the stagnation radius is located in the central
region of the cluster ($r_{\rm st} \lesssim r_0$) resulting in a low accretion
rate (LAR), while at low temperatures ($c_{\rm s} / \sigma < 1$) the stagnation
radius is far from the centre ($r_{\rm st} \gg r_0$) and the accretion rate is
high (HAR). Note that the accretion rate differs by almost three orders of
magnitude between the LAR and the HAR. Fig.~\ref{vstemp_adim} shows another
characteristic differentiating the HAR from the LAR. In the former, the
stagnation radius and the accretion rate increase with the cluster concentration
$c_{\rm GC}$.

\begin{figure}
\vskip 0.7cm
 \includegraphics[width=0.95\columnwidth]{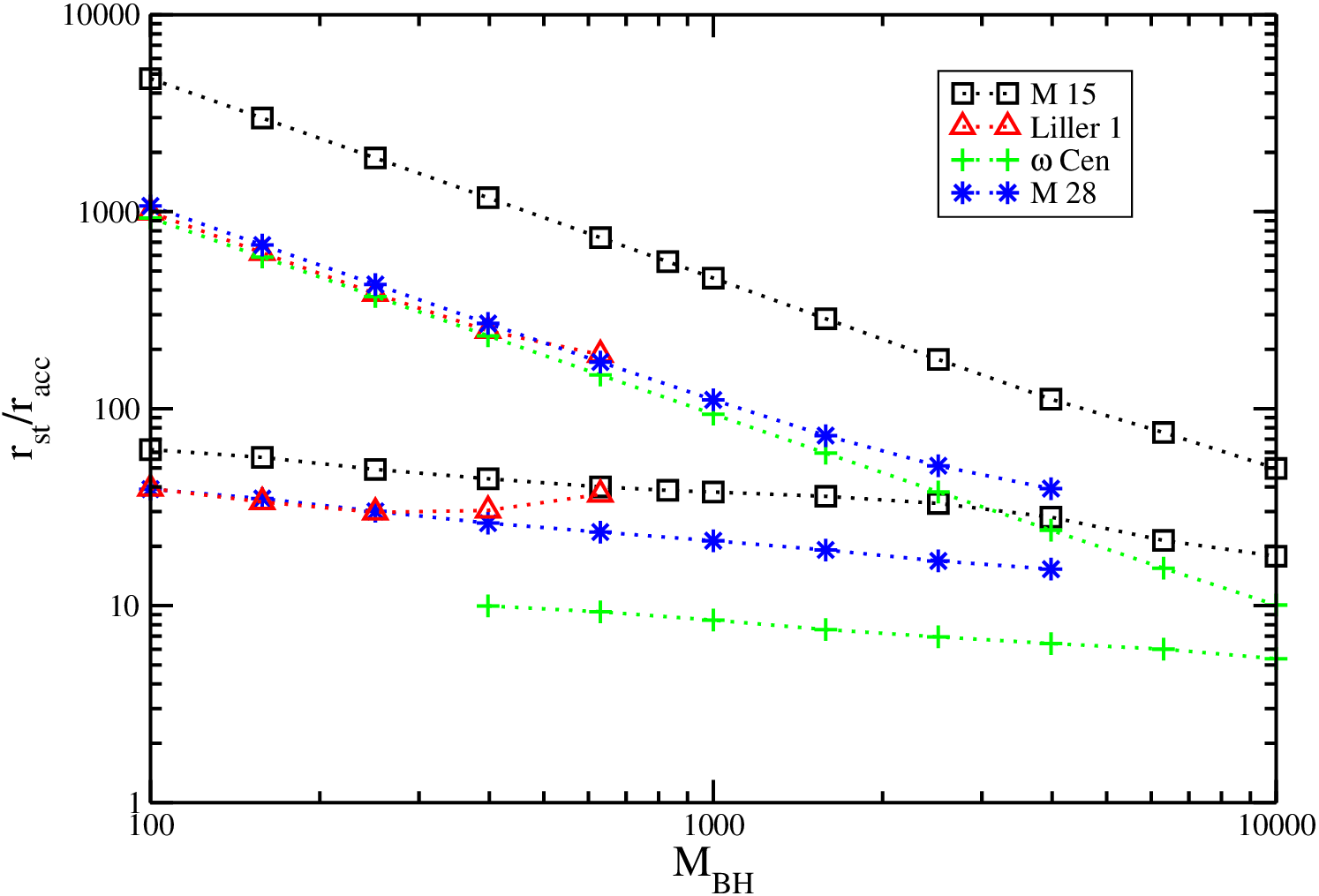}
\caption{Ratio between the stagnation radius and the accretion radius of black hole. In
the LAR regime the stagnation radius behaves like the accretion radius.}
 \label{cociente}
\end{figure}

Fig.~\ref{vsmasa_adim} shows the dependence of the stagnation radius and
accretion rate on the non-dimensional IMBH mass, at fixed non-dimensional
temperature. Two temperatures were explored, one resulting in the LAR and the
other in the HAR. Both the stagnation radius and the accretion rate are almost
independent of $\Omega_{\rm BH}$ in the HAR regime. This happens because far from the
centre the IMBH has no influence on the stellar distribution, and the stellar
mass increases very slowly. On the other hand, in the LAR regime there is a strong
dependence on the IMBH mass, because the stagnation radius is near the
sphere of influence where the stellar distribution becomes dominated by the
IMBH. Note that in this regime, the accretion rate varies also with the cluster
concentration, decreasing as the latter increases. In Fig.~\ref{cociente} we
show the ratio between the stagnation radius and the accretion radius defined
as $r_{\rm acc} = GM_{\rm BH}/c_{\rm s}^2$. It can be seen that in the HAR regime, since the stagnation
radius remains independent of the black hole mass, the ratio $r_{\rm st}/r_{\rm acc}$
decreases with increasing $M_{\rm BH}$. In the LAR regime, this ratio
remains almost constant implying that $\dot{M}$ 
scales as $\sim M_{\rm BH}^2$ like in the traditional Bondi-Hoyle model. However, the
ratio is 1--2 orders of magnitude greater than unity, and hence the accretion
rate at a fixed IMBH mass is much higher in our models than in the Bondi-Hoyle
case.

\subsection{Milky Way globular clusters}

In the previous section, we have shown that in some cases there is a dependence
of the accretion rate on the cluster properties. Here we explore this dependence
by applying the scheme detailed in Sect.~\ref{modelo} to the Milky Way GCs with
well determined parameters \citep[listed in the catalogue of][]{Harris1996}, to
construct flow models for different temperatures and IMBH masses. It is 
worth pointing out that the results from Sect.~\ref{modelo} extend to all
the clusters in the catalogue and the same reasoning can be applied to them. For each
model, the accretion rate was computed using the same value of $\alpha$ as in
the previous section. In some cases, the models were discarded because the IMBH
mass was not much lower than the cluster mass, while in others no stagnation
radius was found for the hottest temperatures (the ICM escapes completely as a
wind). We searched for correlations between the accretion rate and the
properties of the cluster, defined by their scaling parameters and masses. We
found no trend for the accretion rate $\dot{M}$ with $r_0$ or $\rho_0$, but a
clear correlation with the cluster mass $M_{\rm GC}$ and velocity dispersion
parameter $\sigma$.

\begin{figure*}
\vskip 0.7cm
 \includegraphics[width=0.95\columnwidth]{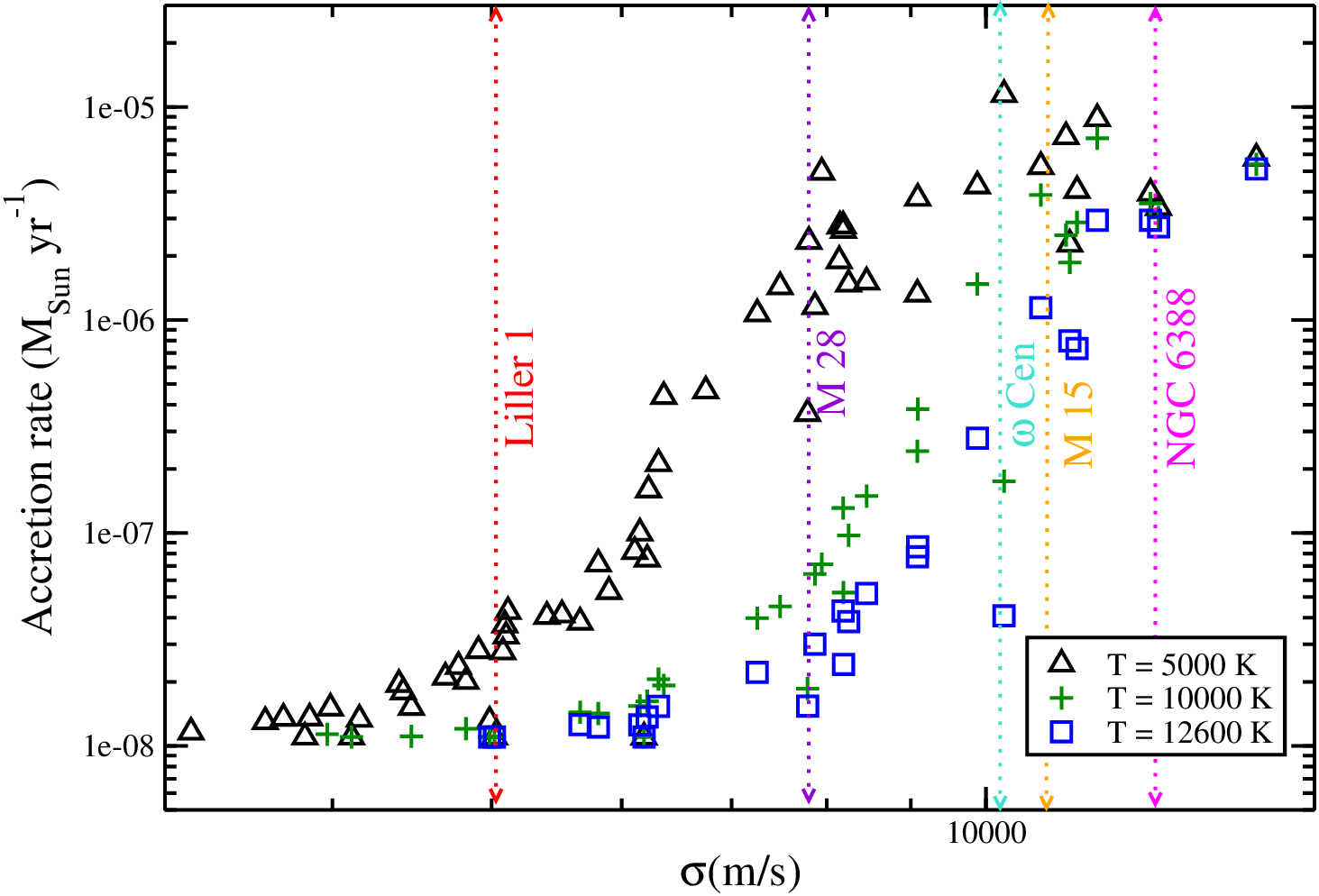}
\hskip 1cm
 \includegraphics[width=0.95\columnwidth]{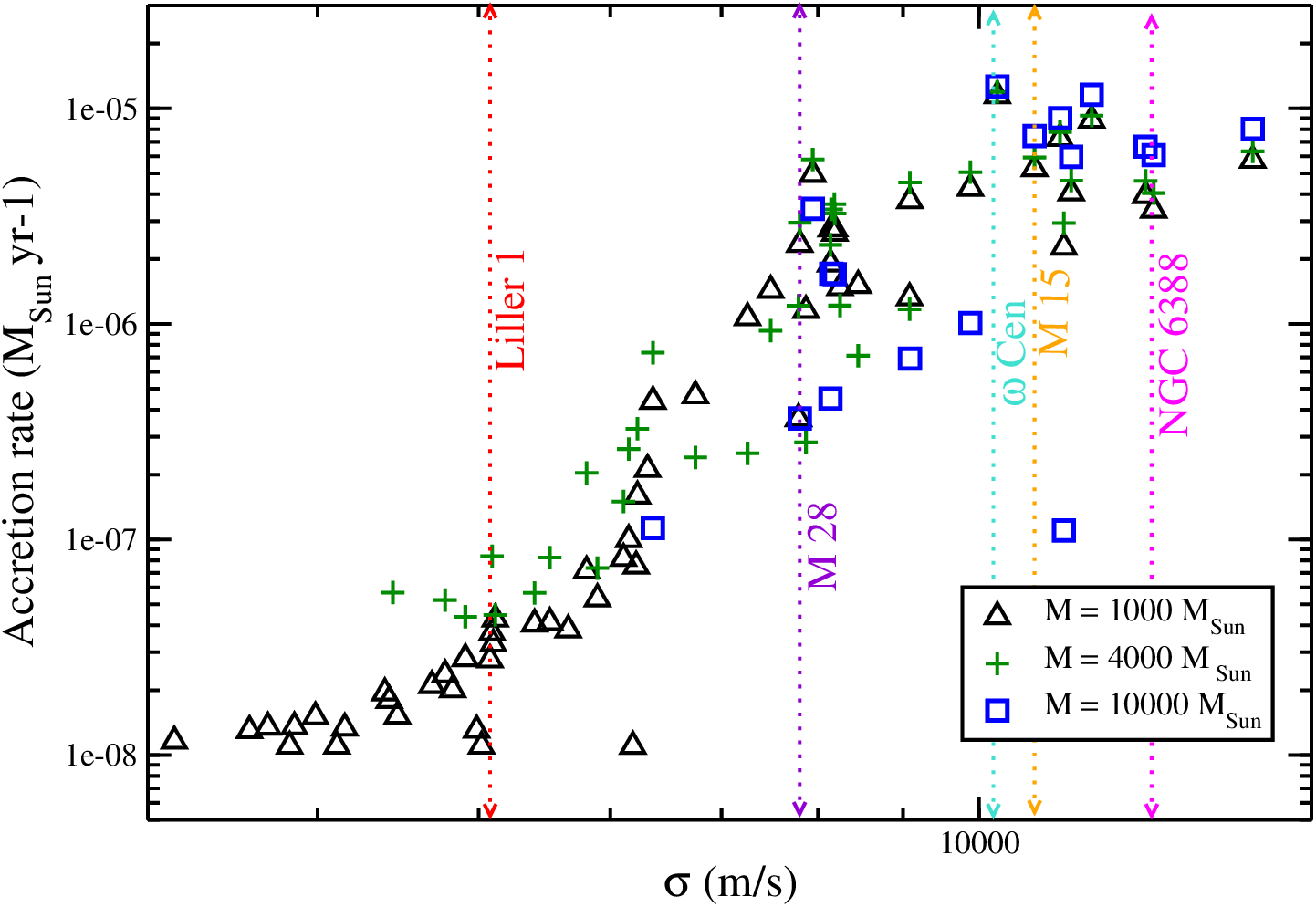}
\caption{Accretion rate vs globular cluster velocity dispersion ($\sigma$) for three different temperatures(left panel): 5000K 
(triangles), 10000K (plus signs) and 12300 K (squares). Accretion rate vs globular cluster velocity dispersion 
for three different black hole masses (right panel): 1000$M_{\odot}$ (triangles) , 4000$M_{\odot}$ (plus signs) and 
10000$M_{\odot}$ (squares). There is 
a nearly constant constant value of $\dot{M}$ for those clusters with low $\sigma$. The clusters located in the right
end of the curve are those clusters accreting at a HAR regime. The dotted lines indicate the $\sigma$ values 
corresponding to the clusters from Sect.~\ref{modelo} and the particular case of NGC 6388 discussed in 
Sect.~\ref{discusion}.  }
 \label{acc_vssigma}
\end{figure*}

We show in Fig.~\ref{acc_vssigma} the dependence between $\dot{M}$ and $\sigma$
for a conservative value of the IMBH mass, $M_{\rm BH} = 1000 M_\odot$ and three
different temperatures $T = 5000,\, 10000,\, 12600\ {\rm K}$ (left panel), and
for a fixed gas temperature ($T = 5000 {\rm K}$) and different IMBH masses
$M_{\rm BH} = 1000,\, 4000,\, 10000M_\odot$ (right panel). The clear trend observed
in the left panel is understood in terms of the results of the previous section.
For a fixed value of $T$, an increase in $\sigma$ implies a decrease in
$c_{\rm s} / \sigma$, changing the flow from the LAR upwards to the HAR in the
curves defined in Figure \ref{vstemp_adim}. This explains why the GCs with
higher $\sigma$ are more likely to be found in a HAR regime. The change in the
location of the curve as the temperature increases is also consistent with
this interpretation. The dispersion of this correlation is in part due to the
different concentrations of the GCs, and in part to the fact that $\dot{M}$
scales as $\rho_0 r_0^3$, which is different for each cluster. The right panel
of Fig.~\ref{acc_vssigma} shows that the effect of the IMBH mass for a fixed
temperature is small.

\begin{figure*}
\vskip 0.7cm
 \includegraphics[width=0.95\columnwidth]{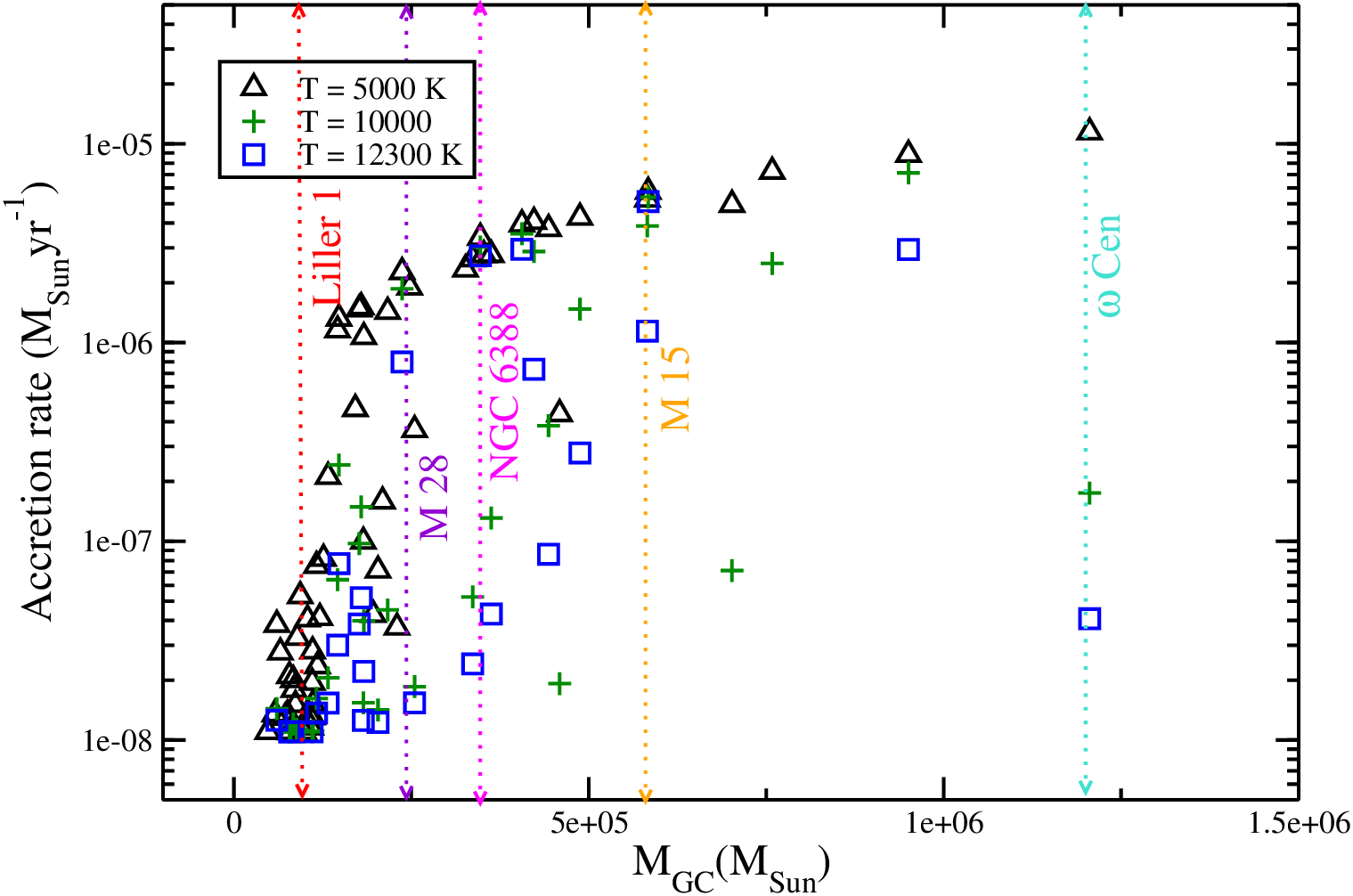}
\hskip 1cm
\includegraphics[width=0.95\columnwidth]{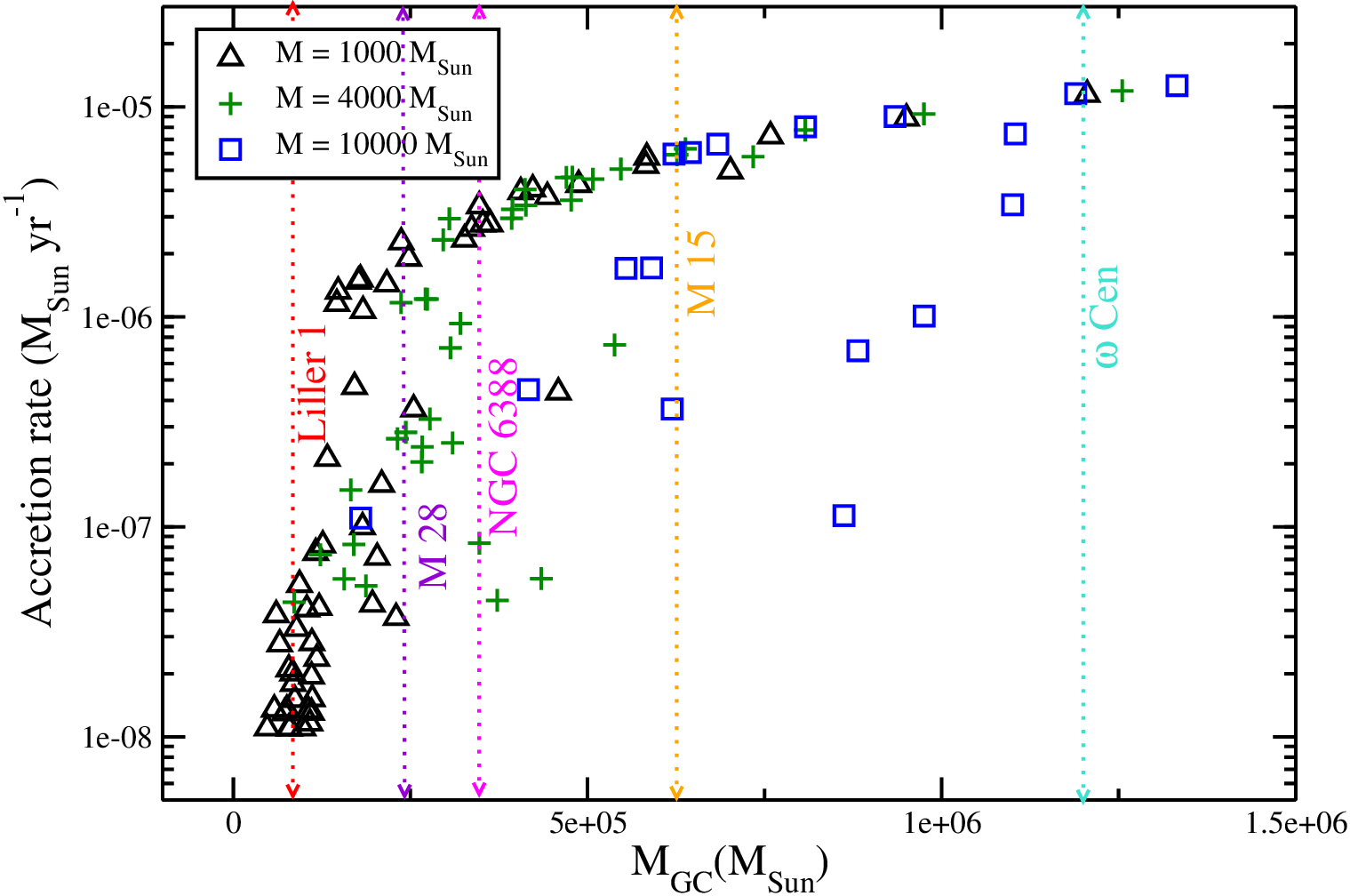}
\caption{Accretion rate vs globular cluster mass for three different temperatures (left panel): 5000K (triangles), 10000K 
(plus signs) and 12300 K (squares). Accretion rate vs globular cluster mass for three differente black hole masses  (right panel):
1000$M_{\odot}$ (triangles) , 4000$M_{\odot}$ (plus signs) and 10000$M_{\odot}$ (squares). A soft trend
can be observed, although there is a wide spread of the data. The dotted lines indicate the $M_{\rm GC}$ values 
corresponding to the clusters from Sect.~\ref{modelo} and the particular case of NGC 6388 discussed in 
Sect.~\ref{discusion}.} 
 \label{acc_vsmasa}
\end{figure*}

A trend of increasing accretion rate with the cluster mass can also be seen in
Fig.~\ref{acc_vsmasa} for high accretion rates. This is explained by the fact
that these clusters are accreting in the HAR regime, for which the stagnation radius is
well outside the cluster core and hence encloses almost the whole cluster
stellar mass. As $\dot{M}$ is proportional to the stellar mass inside
$r_{\rm st}$, the correlation with $M_{\rm GC}$ arises. In other words, most of the
mass ejected by the stars in the cluster is accreted by the IMBH. As the cluster
is more massive, more mass is available to be accreted, which explains the clear
trend with $M_{\rm GC}$ in the HAR regime. Note that this trend is independent of
the temperature and IMBH mass (as far as the accretion proceeds in the HAR
regime), and it is very tight because both $\dot{M}$ and $M_{\rm GC}$ scale with
the same combination of parameters of the cluster ($\rho_0 r_0^3$), reducing the
dispersion due to the scaling from non-dimensional to physical variables. For
clusters in the LAR regime there is no clear trend, as the small stagnation
radius decouples the accretion rate from the cluster mass. According to the
results of this section, the clusters with large values of $\sigma$ and high
$M_{\rm GC}$ are most likely to be in a HAR regime, and are therefore the best
candidates to perform detections of the accretion onto an IMBH. 

\subsection{Differences with the Bondi-Hoyle model}
\label{difference}

\begin{figure}
\vskip 0.7cm
 \includegraphics[width=0.95\columnwidth]{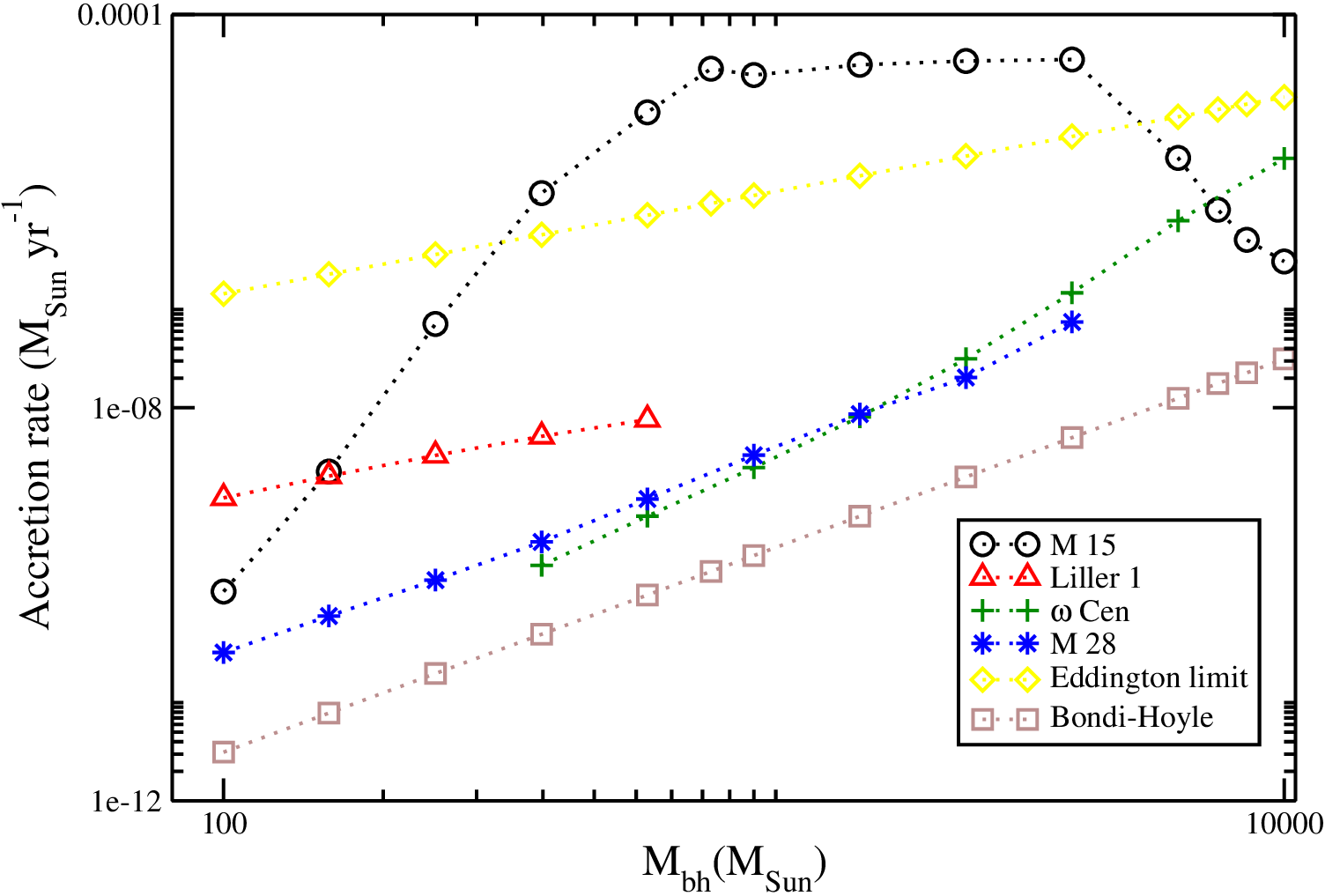}
\caption{Comparison between our accretion rate and Bondi-Hoyle accretion rate. We also show 
the Eddington accretion rate for a reference.}
 \label{comparacion}
\end{figure}

The model presented in this paper differs from that of \cite{Bondi1944} in some
very fundamental aspects. Our model takes into account the gravitational
potential of the cluster, which differs from cluster to cluster, and also
includes the constant injection of gas into the ICM by the red giants of the
cluster, hence the black hole is no longer accreting from a static and infinite
medium. These two features of our model produce a very important consequence,
which is its main difference with the Bondi-Hoyle result: the accretion rate not
only depends on the black hole mass and gas temperature (or sound speed), but
also on the cluster properties. Figure \ref{comparacion} shows the comparison
between our model (for different clusters; all of them at $T = 10000\ {\rm K}$),
the Bondi-Hoyle accretion rate, and the Eddington rate
$\dot{M}_{\rm Edd} = L_{\rm Edd}/c^2$, where $c$ is the speed of light and
$L_{\rm Edd} = 1.26 \times 10^{38} (M_{\rm BH} / M_\odot)\ {\rm erg\ s}^{-1}$ is the
Eddington luminosity of the IMBH. The Bondi-Hoyle accretion rate was calculated
as $\dot{M}_{\rm BH} = 4 \pi G^2 M_{BH}^2 \rho_{\rm a} c_{\rm s}^{-3}$, where
$\rho_{\rm a}$ is the ambient gas density. We used the value
$\rho_{\rm a} = 0.2\ {\rm cm}^{-3}$ \citep{Freire2001}, which is the same value
used by other authors \citep[e.g.,][]{Maccarone2004} to compute IMBH X-ray
luminosities. To make the results of our models comparable to those obtained
with the Bondi-Hoyle scenario, for each of our models we have chosen the
fractional mass loss rate of the stars $\alpha$ so that the density at the
stagnation radius matches $\rho_{\rm a}$. This choice is justified because it
makes both models to have the same density at the point where the fluid is at
rest. The values of $\alpha$ obtained are in the range
$10^{-14}-10^{-11}\ {\rm yr}^{-1}$, in rough agreement with (but somewhat lower
than) the few observational estimations of this parameter \citep{Scott1975, Priestley2011}.
 
It can be seen from Fig.~\ref{comparacion} that the accretion rate in the HAR 
regime 
(such as that of NGC~7078 in this plot) no longer scales as $M_{\rm BH}^2$, and
that it is cluster-dependent. However, in the LAR regime the behaviour is similar to
that of Bondi-Hoyle models, although the absolute value of $\dot{M}$ is one
order of magnitude higher. Another interesting result seen in this figure in
that in the HAR regime, IMBHs reach accretion rates as high as the Eddington values
(assuming a 100\% efficient conversion of gravitational energy into radiation).
This result is due to the development of a cluster-wide flow in the HAR regime,
which carries matter from the outer regions onto the IMBH to produce a high
accretion rate. It cannot be reproduced by the Bondi-Hoyle model unless a very
high density or very low temperature are assumed. This result is important for
the explanation of the emission of ULXs, as will be discused in the next
section. It is worth pointing out that these flows with super-Eddington
accretion rates do not imply super-Eddington luminosities as the radiation
efficiency is usually much lower that 1. Hence, these flows can still proceed
without being stopped by radiation pressure.
 
\section[]{Discussion}
\label{discusion}

In this work, we developed a simple model for the accretion of the ICM of a
globular cluster onto an IMBH at its center. Bearing in mind that the detection
of ongoing accretion would be a strong piece of evidence for the existence of
these elusive black holes, our aim was to refine the predictions of the
expected accretion rate, improving on the (usually assumed) classical
Bondi-Hoyle accretion. In particular, we explored the consequences of including
the effect of the gravitational field of the cluster and the fluid sources on
the flow. Our models aims at assessing the influence of the cluster as a whole
on the accretion flow, which was suggested to be important in the case of
cosmological fluids such as dark matter by \citet{Pepe2012}.

We made several assumptions in our model, namely that the flow is stationary,
spherically symmetric, and isothermal. While the first two can be regarded as
working hypotheses made to simplify the computations, the latter can be a
good approximation for the whole cluster \citep{Scott1975}. However, it may
break down very near the black hole, where the transformation of gravitational
into thermal energy is higher and may heat the gas in timescales shorter than
the cooling one. Although our model could not describe the dynamics of the flow
in these spatial scales, it is still useful to assess the existence of
cluster-wide flows, whose dynamics is governed by phenomena occurring at much
larger spatial scales, and estimates the order of magnitude of the accretion rate
produced by these flows. To analyse the details of the flow very near the black
hole, we are currently developing more complex models which include several
heating and cooling mechanisms.

Our main results are the following. First, large scale flows can form in
globular clusters with IMBHs, due to the influence of the gravitational field
of the cluster and the mass loss of the cluster stars. These flows show a
stagnation radius at which the velocity is null. Outside this radius, the ICM
escapes from the cluster as a wind, as the energy of the flow allows it to
overcome the cluster potential well. Inside the stagnation radius the ICM is
retained by the potential well, and an accretion flow onto the IMBH develops. As
our models are stationary, the accretion rate onto the IMBH is determined by the
stagnation radius and the mass-loss rate of the stars.

The accretion rate onto the IMBH in a given cluster depends mainly on
the ratio of the sound speed in the ICM to the stellar velocity dispersion
parameter. This is a consequence of the energetics of the flow, and the same
effect observed by \citet{Pepe2012} for dark matter. The sound speed
(related to the temperature of the flow) measures the internal energy of the
ICM, while the stellar velocity dispersion parameter indicates the strength of
the cluster gravitational field. As the ratio of these two magnitudes increases,
the ratio of the internal to gravitational energy of the flow increases as well,
making easier for the ICM to escape from the cluster. Therefore the stagnation
radius decreases, the wind becomes stronger and the accretion flow  weaker,
diminishing the accretion rate.

The stellar mass distribution of globular clusters has also important effects
on the flow. These clusters show a large concentration of their mass within a
few core radii of their centres, while the potential well of the cluster
extends to far beyond (typically to tidal radii one or two order of magnitudes
larger than the core radius). This translates into a steep increase of the
accretion rate when the temperature is low enough for the stagnation radius to
reach the core radius, as most of the cluster stellar mass (and hence most of
the stellar mass loss) is enclosed within the latter. This steep increase
separates two accretion regimes with different properties, one with a high
accretion rate at low temperatures (HAR), and the other with a low accretion
rate at higher temperatures (LAR).

In the HAR regime, for a fixed temperature the accretion rate is high, proportional to
the cluster mass, and almost independent of the IMBH mass, as far as the IMBH is
not massive enough to severely change the whole cluster mass distribution
(however, if this were the case, strong signatures of the presence of the black
hole should be found in the stellar distribution and dynamics). The dependence
of the accretion rate on the cluster mass instead on the IMBH mass is due to the
fact that in the HAR regime, the stagnation radius is in the outer region of the GC,
and the IMBH accretes a major fraction of the mass lost by the stars, which is
proportional to the cluster mass. Note the difference with the classical
Bondi-Hoyle model, in which the accretion rate scales as the square of the black
hole mass.

The LAR regime is qualitatively different, showing accretion rates several orders of
magnitude lower, which depend strongly on the IMBH mass. The accretion rates in
the LAR regime are still one order of magnitude greater that the Bondi-Hoyle
prediction for similar boundary conditions, and depend on the IMBH mass because
the stagnation radius is near or inside the IMBH sphere of influence. This
dependence varies from cluster to cluster due to the different stellar-mass
profiles of the clusters, but on average it mimicks the $M_{\rm BH}^2$ dependence
of the Bondi-Hoyle model.

Another strong assumption of our models is that the IMBH is at rest at the
centre of the cluster. This approximation holds because the mass of the IMBH
is far greater that the mass of the stars in the cluster. Assuming that the
stars in the cluster have a mean mass of $\sim 0.5 M_\odot$ and a mean velocity
of the order of $\sigma$, energy equipartition implies that the IMBH would have
a velocity of the order of $\sigma / 2 \sqrt{M_{\rm BH} / M_\odot}$. At these
velocities, and taking into account that $c_{\rm s} / \sigma \sim 1$ in our
models, the typical correction to the accretion rate due to the motion of
the IMBH with respect to the flow \citep{Hoyle1939} would be less than 1\%.
Hence, neglecting this effect is justified for these models.

The application of our model to the Milky Way globular clusters has shown that
the higher accretion rates are predicted for those cluster with the largest
values of the velocity dispersion parameter $\sigma$. However, the higher the
temperature of the gas, the higher the minimum value of $\sigma$ for GCs
accreting at high rates. This result suggests that the signatures of the
accretion onto IMBHs must be searched for in globular clusters with high
velocity dispersion parameters and low gas temperatures. All the clusters that
were observationally proven so far for the existence of IMBHs satisfy the first
criterion. The satisfaction of the criterion on gas temperature is difficult to
assess, as measurements of the ICM thermodynamical state are still rare
\citep[e.g.,][]{Freire2001}. As pointed out by \citet{Scott1975}, the
temperature of the flow depends mainly on that of the radiation field to which
it is exposed, hence a proxy for the former would be the number of UV sources
within the cluster, such as blue horizontal-branch stars or blue stragglers.
These sources would heat the gas, preventing the accretion flow to develop out
to large scales, and hence decreasing the accretion rate. Interestingly,
\citet{Miocchi2007} argues that extreme horizontal-branch stars with strong UV
fluxes might be the result of the stripping of normal stars passing near the
IMBH. If this is indeed the case, the presence of the IMBH would help in heating
the flow, decreasing the accretion rate.

From the results of our model we can estimate the X-ray luminosity $L_{\rm X}$
due to the accretion process onto the IMBH. However, in our model the luminosity
depends linearly on two poorly constrained parameters, the fractional mass loss
rate of the stars $\alpha$ and the accretion efficiency of the flow,
$\epsilon = L_{\rm X} / \dot{M} c^2$. If we adopt the standard values,
$\epsilon = 0.1$ for accreting stellar-mass black holes and
$\alpha = 10^{-11}\ {\rm yr}^{-1}$, the luminosities produced by the accretion
flows of our models would be in the range $10^{37}-10^{41}\ {\rm erg\ s}^{-1}$.
The high end of this range is in good agreement with the luminosities of ULXs.
If, as claimed in a few cases, there are ULXs positionally coincident with extragalactic GCs
\citep[and references therein]{Angelini2001,Maccarone2011}, our models suggest that these systems may contain IMBHs
accreting in the HAR regime at the centres of GCs, with standard accretion
efficiencies.

In the case of NGC~6388 our models predict (based on the estimation of 
the accretion rate and assuming the efficiencies discussed above)
X-ray luminosities in the range
$10^{38-40}\ {\rm erg\ s}^{-1}$, at least 5 orders of magnitude higher than the observed value of
$L_{\rm X,\ NGC6388} = 2.7 \times 10^{33}\ {\rm erg\ s}^{-1}$. However, the prediction
can be reconciled with observations if either $\alpha$ or $\epsilon$ are lower.
\citet{Nucita2008} reaches the same conclusion about the efficiency.
It is worth mentioning that different theoretical models
have been developed for accretion flows with very low efficiencies, such as
advection-dominated accretion flows \citep[ADAFs,][]{narayan1994}, jet-dominated accretion flows \citep[JDAFs,][]{fender2003}, and low-radiative efficiency
accretion flows \citep[LRAFs, ][]{quataert2001}. For example, in the case of 
ADAFs the efficiency scales as $\epsilon = \dot{M}/\dot{M}_{\rm Edd}$ for
$\dot{M} < 0.1 \dot{M}_{\rm Edd}$, giving in our case estimations of the
luminosity 2--3 orders of magnitude lower, approaching the observational
limits. On the other hand, the value of $\alpha$ is highly uncertain,
relying in theoretical models and with few observational constraints
\citep{Scott1975, McDonald2011, Priestley2011}. Given that
the low luminosity limits of our predicted range correspond to low IMBH masses,
we conclude that such an IMBH accreting in the LAR is a possible explanation for
the central engine of the central X-ray source in NGC~6388. The case of G1 is
slightly different, as this cluster is so massive that the luminosity in the
HAR is $L_{\rm X} \sim 10^{42}\ {\rm erg\ s}^{-1}$, and the development of a LAR
requires temperatures of several times $10000\ {\rm K}$. As the observed
luminosity is only  $L_{\rm X} = 2 \times 10^{36}\ {\rm erg\ s}^{-1}$, it is
difficult to reconcile it with the predictions, unless the G1 ICM is strongly
heated, the accretion efficiency is extremely low, or its stars lose mass at
very low rates. The Galactic GCs with undetected central X-ray sources, with
measured upper limits of $10^{31-32}\ {\rm erg\ s}^{-1}$ for their X-ray luminosities are in the
same case.

The models presented in this work explore the consequences of taking into
account the gravitational field of the GC and the mass loss by stars in the
computation of the accretion rate onto an IMBH in the centre of a GC. Although
they are very simple, they allow us to get some insight in the development of
cluster-wide flows that may feed these compact objects. Many improvements can be
made on the models, such as relaxing the isothermal hypothesis to include the
detailed physics of gas heating and cooling, or including the radiation pressure
on the flow due to the X-ray emission of the flow itself. At present we are
working in these tasks, which require a more complex numerical approach to the
problem. We expect that an improved physical model that gives more precise
predictions on the accretion rates and luminosities, together with detailed
stellar evolution models that predict stellar mass loss rates, deeper X-ray
measurements and ICM observations, we would be able to determine the existence
or not of these elusive compact objects and put limits on their masses.

\section*{Acknowledgments}
We thank the anonymous referee for her/his comments and suggestions, that
really helped to improve 
our manuscript. We acknowledge financial support by Argentine ANPCyT 
through grant PICT-2007-0848.

%
%
%
%
%

%
%
%
%

%

\label{lastpage}

\end{document}